\documentclass[12pt,preprint]{aastex}

\shorttitle{Constraints on Cardassian Scenario}
\shortauthors{Zong-Hong Zhu and Masa-Katsu Fujimoto}

\begin{document}

\title{
	Constraints on Cardassian Scenario from 
	the Expansion Turnaround Redshift and the Sunyaev-Zeldovich/X-ray Data
	}

\author{
        Zong-Hong Zhu
        and
        Masa-Katsu Fujimoto
        }

\affil{
        National Astronomical Observatory,
                2-21-1, Osawa, Mitaka, Tokyo 181-8588, Japan\\
        zong-hong.zhu@nao.ac.jp,
        fujimoto.masa-katsu@nao.ac.jp
      }

\begin{abstract}
Cosmic acceleration is one of the most remarkable cosmological findings
  of recent years.
Although a dark energy component has usually been invoked as the mechanism 
  for the acceleration, A modification of Friedmann equation from various
  higher dimensional models provides a feasible alternative.
Cardassian expansion is one of these scenarios, in which the universe is
  flat, matter (and radiation) dominated and accelerating but contains no
  dark energy component.
This scenario is fully characterized by $n$, the power index of the so-called
  Cardassian term in the modified Friedmann equation, and $\Omega_m$,
  the matter density parameter of the universe.
In this work, we first consider the constraints on the parameter space from the
  turnaround redshift, $z_{q=0}$, at which the universe switches from 
  deceleration to acceleration.
We show that, for every $\Omega_m$, 
  there exist a unique $n_{\rm peak} (\Omega_m)$, 
  which makes $z_{q=0}$ reach its maximum value,
  $[z_{q=0}]_{\rm max} = \exp\left[1/ (2-3n_{\rm peak})\right] -1$,
  which is unlinearly inverse to $\Omega_m$.
If the acceleration happans earlier than $z_{q=0} = 0.6$, 
  suggested by Type Ia supernovae measurements, we have $\Omega_m < 0.328$
  no matter what the power index is, and moreover,
  for reasonable matter density, $\Omega_m \sim 0.3$, it is found 
  $n \sim (-0.45,0.25)$.
We next test this scenario using the Sunyaev-Zeldovich/X-ray data of 
  a sample of 18 galaxy clusters with $0.14 < z < 0.83$ compiled by 
  Reese et al. (2002).
We determine $n$ and $\Omega_m$, as well as the Hubble constant $H_0$, using
  the $\chi^2$ minimization method.
The best fit to the
  data gives $H_0 = 59.2\, {\rm kms}^{-1}{\rm Mpc}^{-1}$, $n=0.5$ and 
  $\Omega_m=\Omega_b$ ($\Omega_b$ is the baryonic matter density parameter).
However the constraints from the current SZ/X-ray data is weak, though a
  model with lower matter density is prefered.
A certain range of the model parameters is also consistent with the data. 
\end{abstract}


\section{Introduction}

Recent observations of type Ia supernovae by two independent groups,
  the High-z SuperNova Team\footnote{High-z SuperNova Search: 
  http://cfa-www.harvard.edu/cfa/oir/Research/supernova/HighZ.html} 
  (Riess et al. 1998) and the Supernova Cosmology
  Project\footnote{Supernova Cosmology Project: http://www-supernova.lbl.gov} 
  (Perlmutter et al. 1999) suggest that our universe is presently
  undergoing an accelerating expansion.
The highest redshift supernova observed so far, SN 1997ff at $z=1.755$, 
  not only support this accelerating view, but also glimpse the earlier
  decelerating stage of the expansion (Riess et al. 2001).
It seems that determining a convincing mechanism with a solid basis in
  particle physics that explains the accelerating universe is emerging as
  one of the most important challenges in modern cosmology.

It is well known that all known types of matter with positive pressure generate
  attractive forces and decelerate the expansion of the universe --
  conventionally, a deceleration factor is always used to describe the status 
  of the universe's expansion (Sandage 1988).
Given this, the discovery from the high-redshift type Ia supernovae may
  indicates the existence of a new component with fairly negative
  pressure, which is now generally called dark energy.
Coincidently or not, a dark energy component could offset the deficiency of
  a flat universe, favoured by the measurements of the anisotropy of 
  the cosmic microwave background 
	(de Bernardis et al. 2000; 
	Balbi et al. 2000, 
	Durrer et al. 2003; 
	Bennett et al. 2003; 
	Melchiorri and Odman 2003;
	Spergel et al. 2003),
  but with a very subcritical matter density parameter $\Omega_m \sim 0.3$, 
  obtained from dynamical estimates or X-ray and lensing observations of 
  clusters of galaxies(for a recent summary, see Turner 2002).
The simplest possibility for the dark energy component is 
  cosmological constant $\Lambda$ 
	(Weinberg 1989; 
	Carroll et al. 1992; 
	Krauss and Turner 1995; 
	Ostriker and Steinhardt 1995;
	Chiba and Yoshii 1999;
	Futamase and Hamana 1999).
Other candidates for the dark energy include: 
  a decaying vacuum energy density or a time varying $\Lambda$-term 
	(Ozer and Taha 1987; 
	Vishwakarma 2001; 
	Alcaniz and Maia 2003),
  an evolving scalar field 
	(referred to by some as quintessence: 
	Ratra and Peebles 1988; 
	Wetterich 1988;
	Frieman et al. 1995;
	Coble et al. 1997;
	Caldwell et al. 1998; 
	Wang and Lovelace 2001;
	Wang and Garnavich 2001;
	Podarius and Ratra 2001;
	Li, Hao and Liu 2002;
	Weller and Albrecht 2002;
	Li et al. 2002a,b;
	Chen and Ratra 2003;
	Mukherjee et al. 2003),
  the phantom energy, in which the sum of the pressure and energy 
    density is negative
	(Caldwell 2002;
	Hao and Li 2003a,b;
	Dabrowski et al. 2003),
  the so-called ``X-matter", an extra component simply characterized by an
    constant equation of state $p_{\rm x}=\omega\rho_{\rm x}$ (XCDM) 
	(Turner and White 1997; 
	Chiba et al. 1997;
	Zhu 1998, 2000; 
	Zhu, Fujimoto and Tatsumi 2001; 
	Yamamoto and Futamase 2001;
	Sereno 2002;
	Alcaniz, Lima and Cunha 2003;
	Jain et al. 2003;
	Discus and Repko 2003;
	Lima, Cunha and Alcaniz  2003),
  the Chaplygin gas whose equation of state is given by $p= -A/\rho$ where $A$ 
    is a positive constant 
	(Kamenshchik et al. 2001; 
	Bento et al. 2002; 
	Alam et al. 2003;
	Alcaniz, Jain and Dev 2003;
	Dev, Alcaniz and Jain 2003a; 
	Silva and Bertolami 2003;
	Dev, Jain and Alcaniz 2003).
Although a lot of efforts have been made to pin down the amount and nature 
  of the dark energy, it is still far from reaching a convincing machanism
  with solid basics of particle physics for the accelerating universe.

On the other hand, many models have appeared that make use of the very ideas 
  of branes and extra dimensions to obtain an accelerating universe
	(Randall and Sundrum 1999a,b;
	Deffayet, Dvali and Gabadadze 2002;
	Avelino and Martins 2002;
        Alcaniz, Jain and Dev 2002;
        Jain, Dev and Alcaniz 2002).
The basic idea behind these braneworld cosmologies is that our observable
  universe might be a surface or a brane embedded in a higher dimensional bulk
  spacetime in which gravity could spread (Randall 2002).
The bulk gravity see its own curvature term on the brane which accelerates the
  universe without dark energy.
Here we are concerned with the mechanism proposed by Freese and Lewis (2002),
  in which some kind of bulk stress energy changes the form of
  the Friedmann equation as follows
\begin{equation}
\label{eq:ansatz}
H^2 = A\rho + B\rho^n \,\,,
\end{equation}
where $H \equiv \dot R /R$ is the Hubble parameter as a function of cosmic time,
  $R$ is the scale factor of the universe, 
  $\rho$ is the energy density containing only ordinary matter ($\rho_m$) and 
  radiation ($\rho_r$), i.e., $\rho = \rho_m + \rho_r$.
Since at present $\rho_r \ll \rho_m$, $\rho$ can be considered consisting of
  $\rho_m$ only. 
The second term, called Cardassion term,\footnote{The name Cardassian refers
  to a humanoid race in Star Trek whose goal is to take over the universe,
  i.e., accelerated expansion. This race looks strange to us but is made
  completely of matter (Freese 2003).} 
  would drives the universe accelerating at a late epoch 
  when it is dominated (or before or after it is dominated, which depends on
  the power index $n$ being equal to or less or larger than $1/3$, see
  section 3 for more detail discussion),
  without seeking to another unknown dark energy component.
Several authors have explored the agreement of the Cardassian expansion model
  with various observations such as
  the compact radio source angular size versus redshift data
                (Zhu and Fujimoto 2002),
  the cosmic macrowave background anisotropy measurements 
		(Sen and Sen 2003a, 2003b),
  the distant type Ia supernovae data 
		(Zhu and Fujimoto 2003; Wang et al. 2003; Cao 2003;
		Szydlowski and Czaja 2003; Godlowski and Szydlowski 2003),
  optical gravitational lensing surveys
		(Dev, Alcaniz and Jain 2003b)
  and structure formation
		(Multamaki et al. 2003).
In this work, we first constrain the parameters from the turnaround redshift,
  $z_{q=0}$, at which the universe switches from deceleration to acceleration.
Secondly we analyze this scenario with the Sunyaev-Zeldovich (SZ)/X-ray 
  data compiled by Reese et al. (2002), which is so far the largest and
  homogeneous sample.
After providing the basic equations and the angular diameter distance
  formulas relevant to our analysis (section~2), 
  we discuss the constraints on the parameter space of the Cardassian model
  from the turaround redshift $z_{q=0}$ in section~3.
In section~4, we present our SZ/X-ray data analysis to test the scenario. 
Finally, we finish the paper by summarizing our conclusions and discussion
  in section~5.

\section{Basic equations}

We provide here the important equations resulting from the modified Friedmann
  equation, Eq.(1), for more details, see Freese and Lewis (2002).
In the usual Friedmann equation, $B = 0$. To be consistent with the standard
  comological result, one should take $A = 8\pi G/3$.
It is convenient to use the redshift $z_{eq}$, at which the two terms of
  Eq.(\ref{eq:ansatz}) are equal, as the second parameter of the Cardassian
  model.
In this parameterization of ($n, z_{eq}$), it can be shown that
  (Freese and Lewis 2002),
  $B=H_0^2(1+z_{eq})^{3(1-n)}\rho_0^{-n} [1+(1+z_{eq})^{3(1-n)}]^{-1}$,
  where $\rho_0$ is the matter density of the universe at the present time
  and $H_0=100h\,$kms$^{-1}$Mpc$^{-1}$ is the Hubble constant.

Evaluating the ansatz of Eq.(\ref{eq:ansatz}) at the present time,
  we have (Freese and Lewis 2002)
\begin{equation}
\label{eq:H0}
H_0^2 = {8 \pi G \over 3} \rho_0 [1+(1+z_{eq})^{3(1-n)}] .
\end{equation}
Because in the Cardassian model the universe is flat and contains only matter,
  the matter density at present, $\rho_0$,
  should be equal to the `critical density' of this scenario.
From Eq.(\ref{eq:H0}), we have
\begin{equation}
\label{eq:rhoc}
\rho_0 = \rho_{c,{\rm cardassian}} = \rho_{c} \times F(n,z_{eq}), \;\;\;\;\;
  F(n,z_{eq}) = [1+(1+z_{eq})^{3(1-n)}]^{-1}  \;\; ,
\end{equation}
where $\rho_c = 3H_0^2/8\pi G$ is the critical density of the standard 
  Friedmann model.
As expected for a flat universe, the ``density parameter'',
  $\Omega_{m,{\rm cardassian}} \equiv \rho_0/\rho_{c,{\rm cardassian}}$,
  should be equal to $1$.
But we are used to the standard density parameter,
  $\Omega_m \equiv \rho_0/\rho_c$,
  which is defined in terms of the critical density in standard Friedmann
  model.
From equation (3), we have 
  $\Omega_m = \Omega_{\rm CDM} + \Omega_b = F(n, z_{eq})$,
  and hence we could alternatively use ($n$, $\Omega_m$) to fully characterize 
  the Cardassian model.

Now we evaluate the angular diameter distance as a function of redshift
  $z$ as well as the parameters of the model.
Following the notation of Peebles (1993), we define the redshift 
  dependence of the Hubble paramter $H$ as $H(z) = H_0 E(z)$. 
For the ansatz of Eq.(\ref{eq:ansatz}) and a flat universe with only matter
  (baryonic and cold dark matter), we get
\begin{equation}
\label{eq:newE}
E^2(z; n, \Omega_{\rm CDM}, \Omega_b) = 
	\left[ (\Omega_{\rm CDM} + \Omega_b)\times (1+z)^3 +
	(1-\Omega_{\rm CDM} - \Omega_b)\times (1+z)^{3n} \right]
\end{equation}
Then, it is straightforward to show that the angular diameter distance is 
  given by
\begin{equation}
\label{eq:DA}
D^A(z; H_0, n, \Omega_{\rm CDM}, \Omega_b) = 
	{c \over H_0 }{1 \over {1+z}} \int_{0}^{z} {dz^{\prime} 
		\over E(z^{\prime};n,\Omega_{\rm CDM}, \Omega_b)} .
\end{equation}

\section{Constraints from the turnaround redshift from deceleration
	 to acceleration}

The Cardassian scenario is motivated to provide a possible mechanism for the
  acceleration of the universe.
It is natually that one could use the observational constraints on the
  deceleration parameter, $q(z) \equiv -\ddot{R}R / {\dot{R}}^2$, to check
  the feasibility of the model.
In terms of $E(z)$ function, we get the deceleration parameter as a function 
  of redshift as follows (Zhu and Fujimoto 2003)
\begin{equation}
\label{eq:deceleration}
q(z) \equiv -{\ddot{R}R\over {\dot{R}}^2}
     = -1 + {1 \over 2}{{\rm d}\ln E^2(z;n,z_{eq}) \over {\rm d}\ln (1+z)}
\end{equation}
From Eq.(\ref{eq:newE}), we could derive the turnaround redshift at which
  the universe switches from deceleration to acceleration, or in other words
  the redshift at which the deceleration parameter vanishes, which is as
  follows
\begin{equation}
\label{eq:zq=0}
(1+z)_{q=0} = (2-3n)^{1\over{3(1-n)}} (1+z_{eq}) =
  \left[ (2-3n)({1\over \Omega_m} -1) \right]^{1\over {3(1-n)}} \,\,.
\end{equation}
In Figure~1, we plot this redshift as a function the power index of the 
  Cardassian term, $n$, for several values of $\Omega_m$.
While $\Omega_m=0.330\pm 0.035$ is the current optimistic matter density
  (see Turner 2002 for the argument), $\Omega_m=0.2$--$0.4$ is a wider range.
The dotted lines of Figure~1 correspond to the present observational
  constraints on the turnaround redshift (at the $1\sigma$ level),
  $0.6 < z_{q=0} <1.7$,
  from the latest supernova data (Perlmutter et al. 1999; 
   Riess et al. 1998,2001; Turner and Riess 2002; Avelino and Martins 2002).
As Figure~1 shows, for every value of $\Omega_m$, there exists a value for
  the power index of the Cardassian term, $n_{\rm peak}(\Omega_m)$, satisfying
\begin{equation}
\label{eq:npeak}
{1\over {2-3n_{\rm peak}}} \exp\left[{3(1-n_{\rm peak})\over{2-3n_{\rm peak}}}\right] = {1\over \Omega_m} -1   \,\,,
\end{equation} 
which makes the turnaround redshift $z_{q=0}$ reach the maximum value, 
\begin{equation}
\label{eq:zq0max}
[z_{q=0}]_{\rm max} = \exp\left[1/ (2-3n_{\rm peak} (\Omega_m))\right] -1  \;\;.
\end{equation}
This $n_{\rm peak}(\Omega_m)$ -- $[z_{q=0}]_{\rm max}$ relation
  is illustrated in Figure~1 by the thick dotted-dash line.
For example, for $\Omega_m = 0.2$, we have from Eq.(\ref{eq:npeak}) 
  $n_{\rm peak} \sim 0.20$ and hence 
  $[z_{q=0}]_{\rm max} \sim 1.05$, or in other words a universe
  with $\Omega_m = 0.2$ can not switch from deceleration to acceleration 
  at redshift higher than $1.05$, no matter what the power index of 
  the Cardassian term is.
We could restate the constraints in an alternative way: for every value of
  the turnaround redshift $z_{q=0}$, the matter density of the universe must
  satisfy
\begin{equation}
\label{eq:Omegammax}
\Omega_m \leq \left[ \ln(1+z)_{q=0} \cdot {(1+z)_{q=0}}^{[\ln(1+z)_{q=0} +1]/\ln(1+z)_{q=0}} + 1\right]^{-1}  \,\,,
\end{equation} 
where the equal mark comes into existence if and only if the power index of the
  Cardassian term is $n_{\rm peak} =[2\ln(1+z)_{q=0} -1]/[3\ln(1+z)_{q=0}]$.
For example, for $z_{q=0} = 0.6$, as we are considering, 
  we have from Eq.(\ref{eq:Omegammax}) 
  $\Omega_m \leq 0.328$, where the equal sign happans at 
  $n = n_{\rm peak} = -0.04$.
The problem is now apparent: a very low matter density would be always
  necessary for the Cardassian expansion scenario if the turneraround redshift
  is larger than 0.6, 
  especially for the case of $n$ deviating from $n_{\rm peak}$.


In order to illustrate how the turnaround redshift efficiently constrains the
  Cardassian parameter space, we explore the model with different values in
  the $n$--$\Omega_m$ plane.
The results are shown in Figure~2.
The thick solid curve delimits the parameter space of 
  accelerating/decelerating universe as the down-left/up-right area of 
  the curve.
While the present observational constraints on the turnaround redshift
  with $0.6 < z_{q=0} <1.7$ restrict the parameter space into the narrow
  shaded area, the observed matter density of the universe limits them further
  (see the overlap part of the shaded and the cross-hatched areas).
For instance, if the density parameter of the universe takes value around
  $\Omega_m \sim 0.3$, a reasonable value suggested by various observations,
  the power index should be around $n \sim (-0.45,0.25)$ to satisfy
  $z_{q=0} > 0.6$.
Last but not least, we would like to note that, in the accelerating area, 
  a horizontal line with $n=1/3$ will delimit it into two parts, 
  $n>1/3$ or $n<1/3$, corresponding to a universe which switches from
  deceleration to acceleration later or earlier than the moment at which 
  the Cardassian term equals to the conventional term of the Friedmann equation
  respectively.
This can be see from the relation of  equation (7).
For clarity, we do not show it in Figure~2.
Therefore the expansion turnaround redshift is generally not equal to 
  $z_{eq}$, the redshift at which the Cardassian term reach the normal 
  matter density term in Friedmann equation.
Only for $n = 1/3$, the acceleration happans exactly when the Cardassian term
  starts to dominate.
However for $n > 1/3$ ($n < 1/3$), the universe switches from deceleration
  to acceleration after (before) the Cardassian term dominates. 

It is widely believed that the universe only switches from deceleration to
  acceleration recently, which is based on two observational arguments
  (Amendola 2003): (1)acceleration at high redshift might be in contrast
  with the observed large-scale structure, because it makes the gravitational
  instability ineffective; (2)the highest redshift supernova SN1997ff at 
  $z = 1.755$ discovered so far seems to provide a glimpse of the epoch of
  deceleration (Riess et al. 2001; Turner and Riess 2002).
However a cosmological model with a dark energy component strongly coupled
  to dark matter (Amendola 2000,2003) can be consistent with the above two 
  observations but allow acceleration at high redshift, $z_{q=0} \in (1,5)$.
We can see from above analysis that the Cardassian expansion model with 
  reasonable matter density, $\Omega_m \in (0.2,0.4)$, can hardly explain 
  an acceleration happaned earlier than $z_{q=0} =1$.
In this sense, a directly observation of the acceleration might be one of
  the most crucial and efficient tests to discriminate different machanisms
  for acceleration.
There have been so far two proposals for this kind of measurements.
One is to monitor the redshift change of quasars during ten years or so 
  (Loeb 1998).
Another one is to measure the gravitational wave phase of neutron star binaries
  at $z > 1$ for ten years using a decihertz interferometer gravitational wave
  observatory (DECIGO: Seto, Kawamura and Nakamura 2001).
They might provide an efficient test for various acceleration mechanisms.


\section{Constraints from the SZ/X-ray data}

It has long been suggested that a measurement of the thermal SZ
  effect (Sunyaev and Zeldovich 1972) of a cluster of galaxies, combined 
  with X-ray observations, can be used to determine the cluster cosmological 
  angular-diameter distance and hence the Hubble constant $H_0$ and the 
  deceleration parameter $q_0$ 
	(Cavaliere, Danese and de Zotti 1977; 
	Silk and White 1978;
	Birkinshaw 1979).
Benefiting from the improvements to the traditional single-dish
  observations (Myers et al. 1997) and thanks to the new bolometer 
  technology (Holzapfel et al. 1997) and interferometry technique
  (Calstrom, Joy and Grego 1996), several tens of clusters of galaxies have 
  been selected for the SZ measurements and have been used to determine 
  the Hubble constant 
	(for recent summaries, see Birkinshaw 1999; 
	Calstrom, Holder and Reese 2002).
Recently, Reese et al. (2002) published the SZ measurements of a sample
  of 18 galaxy clusters with redshifts ranging from 0.14 to 0.83 observed by
  The Owens Valley Radio Observatory (OVRO) and the Berkeley-Illinois-Maryland
  Association (BIMA) interferometers.
The author used the maximum likelihood joint analysis method to analyze their
  SZ measurements with archival ROSAT X-ray imaging observations, which provide
  the largest homogeneously analyzed sample of the SZ/X-ray clusters with 
  angular diameter distance determinations so far (Reese et al. 2002). 
The database is shown in Figure~3.
Because the redshift range of the cluster sample is comparable to the distant
  type Ia supernovae data compiled by Riess et al. (1998) and Perlmutter et al.
  (1999) that lead to the discovery of accelerating expansion, it provides 
  well an independent cross check of the acceleration mechanism.
We use this sample to give an observational constraint on the Cardassian model 
  parameters, $n$ and $\Omega_m$ (or $\Omega_{\rm CDM}$). 


We determine the model parameters $n$ and $\Omega_{\rm CDM}$ 
  through a $\chi^{2}$ minimization method.
The range of $n$ spans the interval [-1, 1] in steps of 0.01, while the
  range of $\Omega_{\rm CDM}$ spans the interval [0,1-$\Omega_b$] in 101 
  points.
\begin{equation}
\label{eq:chi2}
\chi^{2}(n, H_0, \Omega_{\rm CDM}, \Omega_b) =
  \sum_{i}^{}{\frac{\left[D^A(z_{i}; H_0, \Omega_{\rm CDM}, \Omega_b)
     - {D^A}_{oi}\right]^{2}}{\sigma_{i}^{2}}},
\end{equation}
where $z_i$ are the redshifts of the galaxy clusters and 
  $D^A(z_{i}; H_0, \Omega_{\rm CDM}, \Omega_b)$ are the theoretical predictions
  from equation~(\ref{eq:DA}).
${D^A}_{oi}$ are the angular diameter distances of the cluster sample 
  determined by the SZ/X-ray route.
$\sigma_{i}$ are the symmetric root-mean-square errors of the cluster
  distance measurements (Reese et al. 2002; Allen, Schmidt and Fabian 2002).
The summation is over all the observational data points.

In order to reduce the number of parameters to be determined, 
  we specify the baryonic matter density parameter to be
  $\Omega_b h^2 = 0.0205$ (O'Meara et al. 2001; Allen, Schmidt and Fabian 2002).
In order to make the analysis independent of choice of the Hubble constant,
  we minimize Eq.(\ref{eq:chi2}) for $H_0$, $n$ and $\Omega_{\rm CDM}$,
  simultaneously,
  which gives $H_0=$59.6$\,$km$\,$s$^{-1}\,$Mpc$^{-1}$, $n=0.42$ and 
    $\Omega_{\rm CDM}=0.0$ 
    as the best fit with $\chi^2 = 16.21$ and 15 degrees of freedom (d.o.f).
An inspection of terms of $\chi^2 = 16.21$ for the above best fit shows that
  in this summation a big amount, namely $\sim 5.8$, is given by one cluster
  Abell 370 with redshift 0.374, which should be considered as an outlier.
Following Reese et al. (2002), we also analyze the subsample of 17 galaxy 
  clusters which excludes the outlier, Abell 370.
From the view of statistics, an outlier appears when it is inconsistent with
  the remainder of the data set, and is better to be removed from the sample
  (M\'esz\'aros 2002).
A glimpse of Figure~3, one immediately find that Abell 370 might be an outlier,
  because it is far above the distances expected from the trend given by other
  objects, in another word, it is ``too far''.
Our best fit to the rest 17 clusters occurs for 
  $H_0=$59.2$\,$km$\,$s$^{-1}\,$Mpc$^{-1}$, $n=0.50$ and $\Omega_{\rm CDM}=0.0$
  with $\chi^2 = 10.55$ and 14 d.o.f.
In Figure~4, we show contours of constant likelihood
  (68\% and 95\% C.L.) in the $n$--$\Omega_m$ plane
  for the best fit.
The fitting results for samples with and without Abell 370 are very similar,
  but the $\chi^2$ per d.o.f for the latter reduces significantly and indicates
  that no large statistical errors remain unaccounted for.
The similarity of the fitting results also indicates the robustness of our
  analysis.
The best fits of the Cardassian expansion model to the current SZ/X-ray data
  lead to a universe containing baryonic matter only, leaving no space 
  for the huge amount of dark matter whose existence has been widely
  accepted among the astronomical community.
However, note from Figure~4 that the allowed range for 
  both $n$ and $\Omega_{\rm CDM}$ is reasonable large, showing that 
  the constraints on Cardassian expansion from the considered SZ/X-ray cluster
  sample are not restrictive.
For example, the models with reasonable matter density ($\Omega_m \sim 0.3$)
  but some negative values of $n$ are within $1\sigma$ of the best fit.


\section{Conclusions and discussion}

We have explored the constraints on the parameter space of the Cardassian
  scenario from the turnaround redshift, $z_{q=0}$, at which the universe 
  switches from deceleration to acceleration.
We demonstrated that, for every matter density, $\Omega_m$,
  there exist a unique power index, $n_{\rm peak} (\Omega_m)$,
  which makes $z_{q=0}$ reach its maximum value.  
If the acceleration happans earlier than $z_{q=0} = 0.6$,
  it is found $n \sim (-0.45,0.25)$ for a reasonable matter density, 
  $\Omega_m \sim 0.3$.
It is also found that the Cardassian scenario can hardly explain an 
  acceleration happaned earlier than $z=1$.

We further analyzed the scenario
  using the largest homogeneously sample of SZ/X-ray clusters
  compiled by Reese et al. (2002).
All best fitting results seem to give a universe with very low matter 
  density ($\Omega_m = \Omega_b = 0.0205h^{-2}$).
Our analysis is consistent with previous observational constraints 
	(Sen and Sen 2003a,2003b; 
	Wang et al. 2003;
	Cao 2003;
	Zhu and Fujimoto 2002, 2003).
However at present both the statistical and systematic uncertainties in the
  SZ/X-ray route of distance determination are very large 
	(Birkinshaw 1999; 
	Reese et al. 2002; 
	Calstrom, Holder and Reese 2002).
  which loosen the constraints of the Cardassian expansion scenario very much
  (Figure~4). 
Therefore, the Cardassian expansion scenario with reasonable matter density 
  is also consistent with the current SZ/X-ray data within $1\sigma$.
The projection effect of aspherical clusters modeled with a spherical geometry
 is a large systematic error in the angular diameter distance determinations
  based on the SZ/X-ray route (Hughes and Birkinshaw 1998).
Other large systematic uncertainties are due to the departures form the 
  isothermality, the possibility of clumping and possible point source
  contamination of the SZ observations 
	(Calstrom, Holder and Reese 2002;
	Reese et al. 2002).
It is promising to improve the systematic uncertainties for both SZ and X-ray
  observations within a few years.
The absolute calibration of the SZ observation is now underway to improve
  significantly (Calstrom, Holder and Reese 2002).
The Chandra and XMM-Newton X-ray telescopes not only reduce the uncertainty
  in the X-ray intensity scale, but also are providing temperature profiles
  of galaxy clusters 
	(Markevitch et al. 2000;
	Nevalainen, Markevitch and Forman 2000).
The undergoing or planing large survey of SZ clusters will perhaps provide
  a SZ/X-ray sample of a few hundred clusters with redshift extending to one
  and beyond, which will improve our analysis and pin down the Cardassian
  model parameters more accurately.

\acknowledgements

We would like to thank
   S. W. Allen,
   A. G. Riess,
   and
   R. G. Vishwakarma
   for their helpful discussion.
Our thanks go to the anonymous referee for valuable comments and
   useful suggestions, which improved this work very much.
This work was supported by
   a Grant-in-Aid for Scientific Research on Priority Areas (No.14047219) from
   the Ministry of Education, Culture, Sports, Science and Technology.

\clearpage

\begin{figure}
\plotone{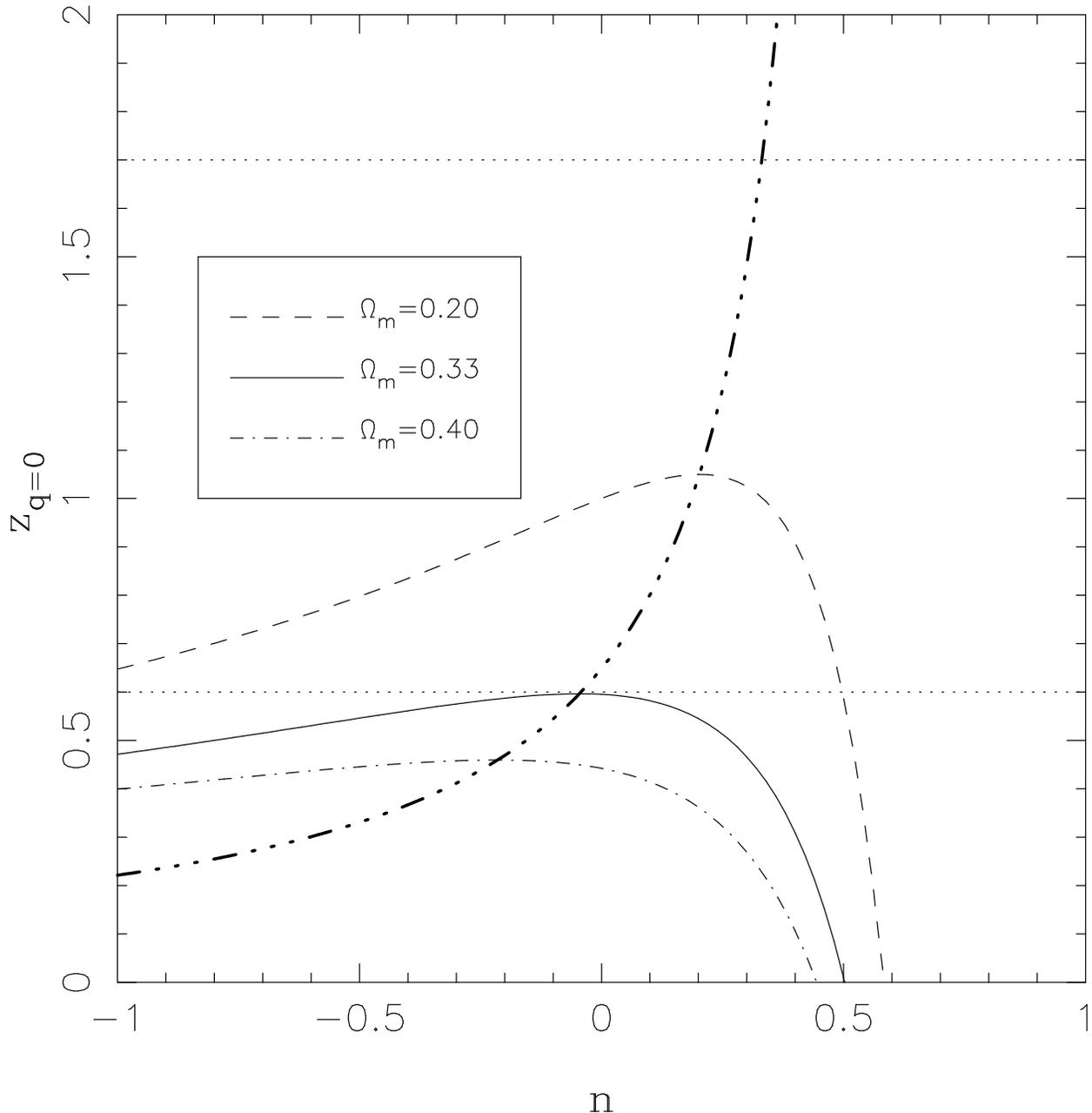}
\figcaption{The turnaround redshift, at which the universe switches from
        deceleration to acceleration, as a function of the power index of
        the Cardassian term, $n$, for three values of the unverse matter
        density, $\Omega_m$.
        While $\Omega_m=0.330$ is the current optimistic matter density,
        $\Omega_m=0.2$ and $0.4$ are the lower and upper bound for a wider
        range respectively.
        The two dotted lines show the present observational constraints with
        $0.6 < z_{q=0} <1.7$ from literatures.
        The thick dotted-dash line depicts the
	$n_{\rm peak}(\Omega_m)$ -- $[z_{q=0}]_{\rm max}$
        relation (see text for details).
        As it shows, a very low matter density is generally necessary for
        the Cardassian model to pass the turnaround redshift test.
        \label{fig:turnaround}
        }
\end{figure}

\clearpage

\begin{figure}
\plotone{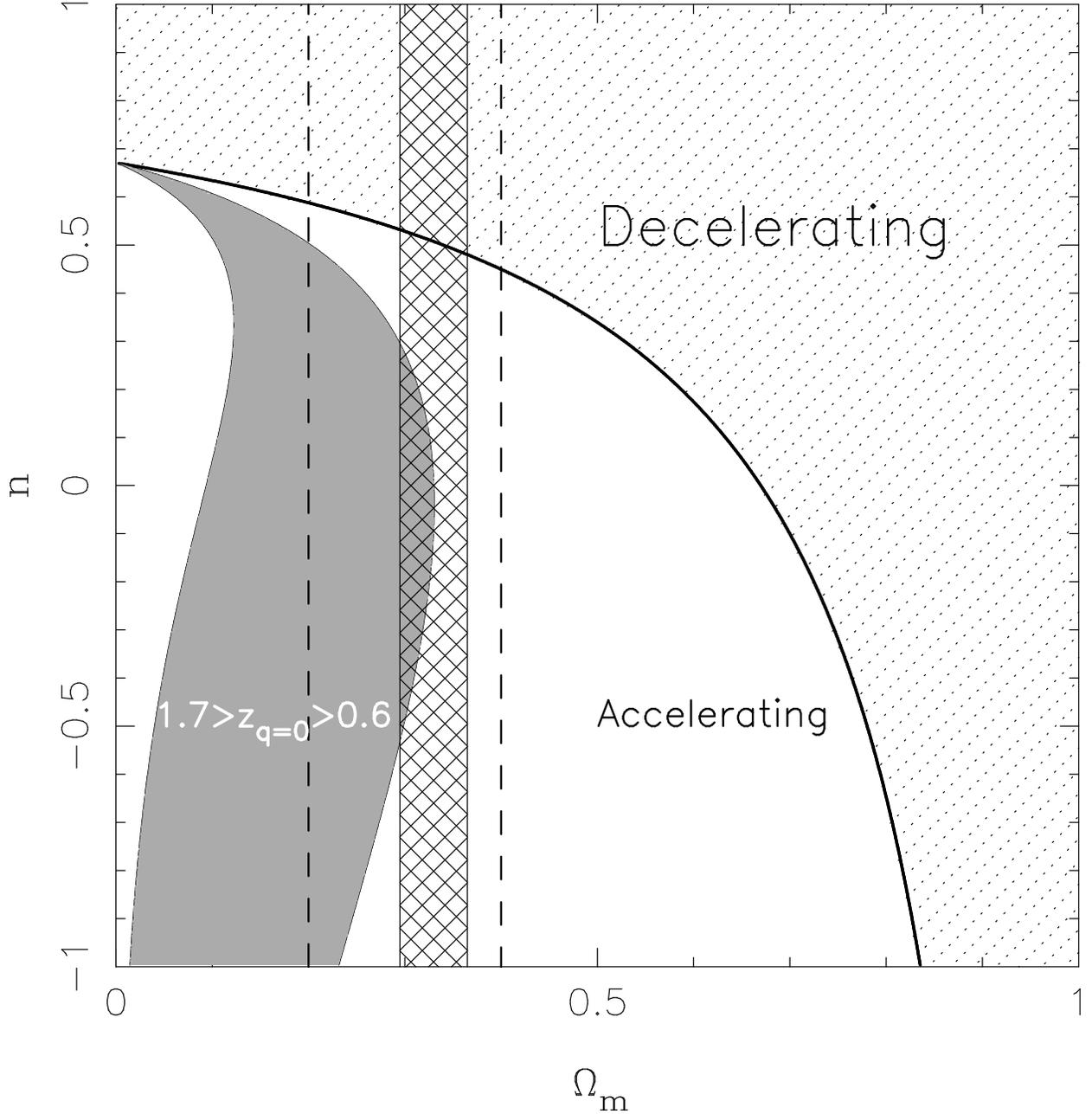}
\figcaption{The constraints on the Cardassian parameter space,
        the $n$--$\Omega_m$ plane,
        from the turnaround redshift of acceleration.
        The thick solid curve, with $z_{q=0} = 0$, is the boundary between
        decelerating (the up and right area of the solid curve) and accelerating
        (the down and left area of the solid curve) models.
        The shaded area corresponds to the Cardassian model parameters that
        can satisfy the present observational constraints with
        $0.6 < z_{q=0} <1.7$ from literatures.
        While the cross-hatched area corresponds to the current optimistic
        matter density, $\Omega_m = 0.330 \pm 0.035$, the two dashed lines
        represent $\Omega_m=0.2$ and $0.4$, the lower and upper bound for
        a wider range, respectively.
        As it shows, although there is some narrow parameter space that could
        satisfy both the turnaround redshift and matter density constraints,
        Cardassian expansion with low matter density is generally
        necessary to be as a possible mechanism for acceleration happaned
        at redshift higher than 0.6.
        \label{fig:parameterspace}
	}
\end{figure}

\clearpage

\begin{figure}
\plotone{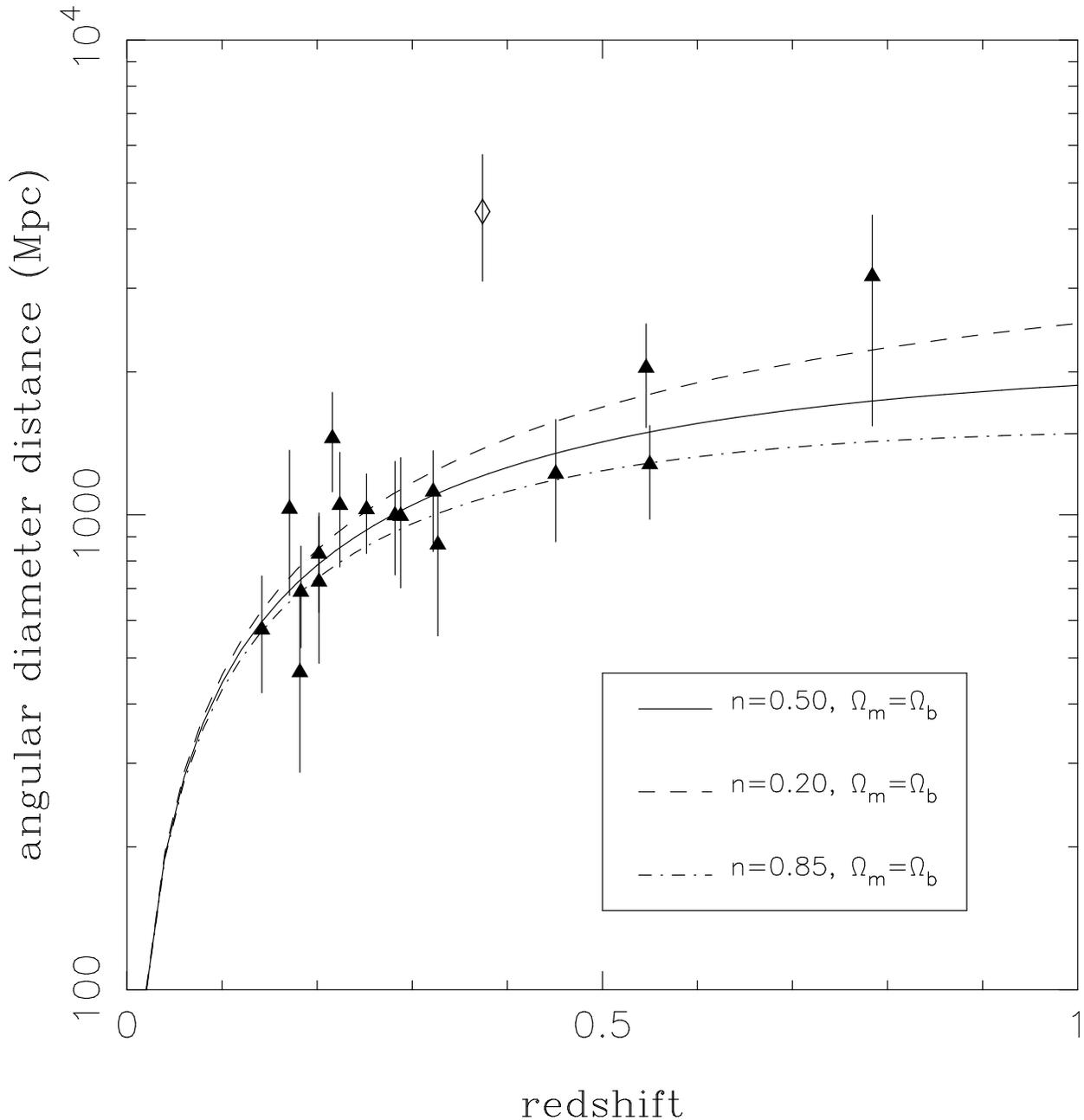}
\figcaption{Diagram of angular diameter distance vs redshift for 18 X-ray
	galaxy clusters compiled by Reese et al. (2002), in which the distances
	are determined by the Sunyaev-Zeldovich/X-ray route.
	The solid curve corresponds to the best fit of the Cardassian expansion
	model to the subsample in which 
	the outlier, cluster Abell 370 marked by the empty diamond, is excluded.
	The values of ($n$, $\Omega_m$) for the other two curves are taken
	from the edge of the allowed regions shown in Figure~4.
	We assume the baryonic matter density to be $\Omega_b = 0.0205h^{-2}$.
	\label{fig:data}
	}
\end{figure}

\clearpage

\begin{figure}
\plotone{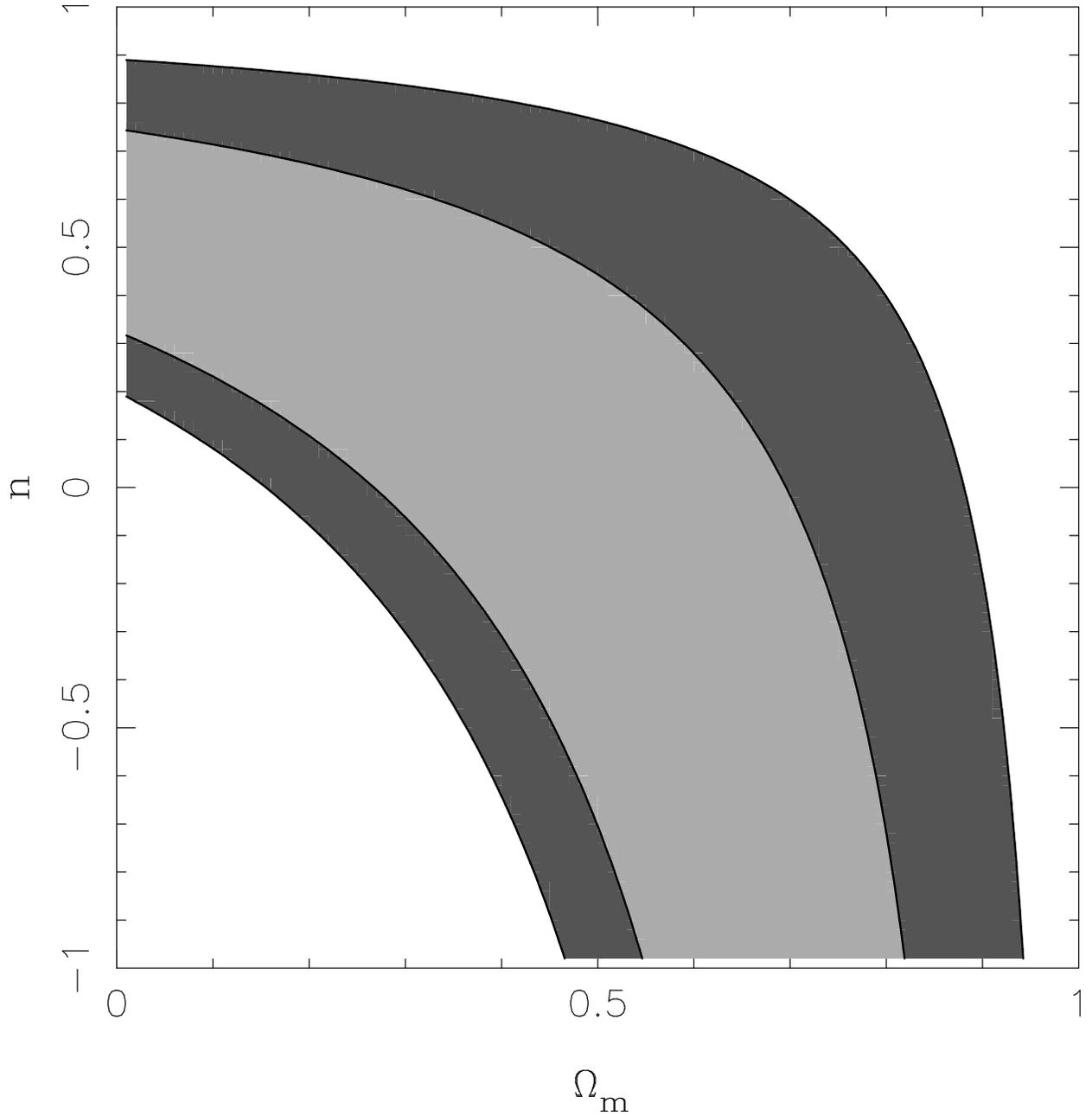}
\figcaption{Confidence region plot of the best fit 
	to the angular diameter distance data of galaxy clusters compiled 
	by Reese et al. (2002), in which the outlier Abell 370 is excluded. --
	see the text for a detailed description of the method.
	The 68\% and 95\% confidence levels in the $n$--$\Omega_m$ plane 
	are shown in lower shaded and lower $+$ darker shaded areas 
	respectively.
	\label{fig:contours}
        }
\end{figure}

\end{document}